\documentclass[12pt]{article}
\usepackage{graphics,graphicx}
\usepackage{amssymb,epsfig,amsmath,euscript,array}
\usepackage{cite}
\usepackage{axodraw}
\usepackage{pstricks}
\usepackage{color}

\makeatletter
\@addtoreset{equation}{section}
\makeatother

\newcounter{multieqs}

\newcommand{\be}{\begin{equation}}
\newcommand{\ee}{\end{equation}}

\newcommand \vev [1] {\langle{#1}\rangle}

\newcommand{\bm}[1]{\mbox{\boldmath $#1$}}

\newcommand{\kslash}{k \!\!\! / }

\newcommand{\lslash}{l \!\! / }
\newcommand{\Pslash}{P \!\!\!\! / }

\newcommand{\islash}{i \!\!\! / }
\newcommand{\jslash}{j \!\!\! / }
\newcommand{\aslash}{a \!\!\! / }
\newcommand{\bslash}{{b \hspace{-6pt} \slash} }

\newcommand{\onslash}{1 \!\!\! / }
\newcommand{\twslash}{2 \!\!\!/ }
\newcommand{\thslash}{3 \!\!\!/ }
\newcommand{\foslash}{4 \!\!\! / }
\newcommand{\fislash}{5 \!\!\! / }

\newcommand{\mslash}{m \!\!\! / }

\def\bd{\begin{document}}
\def\ed{\end{document}}
\def\nn{\nonumber}
\def\bea{\begin{eqnarray}}
\def\eea{\end{eqnarray}}

\def\ab{(ijab)}
\def\ba{(ijba)}
\def\ijab{{\tr}_{+}(\islash\, \jslash\, \aslash \, \bslash)}
\def\ijba{{\tr}_{+}(\islash\, \jslash\, \bslash \, \aslash)}
\def\ijaP{{\tr}_{+}(\islash\, \jslash\, \aslash \, \Pslash)}
\def\ijPLa{{\tr}_{+}(\islash\, \jslash\, \Pslash_L \, \aslash)}
\def\ijaPL{{\tr}_{+}(\islash\, \jslash\, \aslash \, \Pslash_L)}
\def\ijPLza{{\tr}_{+}(\islash\, \jslash\, \Pslash_{L;z} \, \aslash)}
\def\ijaPLz{{\tr}_{+}(\islash\, \jslash\, \aslash \, \Pslash_{L;z})}
\def\ijPa{{\tr}_{+}(\islash\, \jslash\, \Pslash \, \aslash)}
\def\iaPb{{\tr}_{+}(\islash\, \aslash\, \Pslash \, \bslash)}
\def\ibPa{{\tr}_{+}(\islash\, \bslash\, \Pslash \, \aslash)}
\def\ijPmu{{\tr}_{+}(\islash\, \jslash\, \Pslash \, \mu)}
\def\ibmuP{{\tr}_{+}(\islash\, \bslash\, \mu \, \Pslash)}
\def\ibmua{{\tr}_{+}(\islash\, \bslash\, \mu \, \aslash)}
\def\iamub{{\tr}_{+}(\islash\, \aslash\, \mu \, \bslash)}
\def\jaPb{{\tr}_{+}(\jslash\, \aslash\, \Pslash \, \bslash)}
\def\ijmuP{{\tr}_{+}(\islash\, \jslash\, \mu \, \Pslash)}
\def\ijmum{{\tr}_{+}(\islash\, \jslash\, \mu \, \mslash)}
\def\ijmmu{{\tr}_{+}(\islash\, \jslash\, \mslash \, \mu)}
\def\ijmP{{\tr}_{+}(\islash\, \jslash\, \mslash \, \Pslash)}
\def\iabP{{\tr}_{+}(\islash\, \aslash\, \bslash \, \Pslash)}
\def\ijbP{{\tr}_{+}(\islash\, \jslash\, \bslash \, \Pslash)}
\def\jbPa{{\tr}_{+}(\jslash\, \bslash\, \Pslash \, \aslash)}
\def\ijPb{{\tr}_{+}(\islash\, \jslash\, \Pslash \, \bslash)}
\def\jbmua{{\tr}_{+}(\jslash\, \bslash\, \mu \, \aslash)}

\def\loablt{ {\tr}_{+}(\lslash_1\, \aslash \, \bslash\, \lslash_2)}

\def\ijlolt{{\tr}_{+}(\islash\, \jslash\, \lslash_1 \, \lslash_2)}
\def\ijltlo{{\tr}_{+}(\islash\, \jslash\, \lslash_2 \, \lslash_1)}
\def\ibloa{{\tr}_{+}(\islash\, \bslash\, \lslash_1 \, \aslash)}
\def\jaltb{{\tr}_{+}(\jslash\, \aslash\, \lslash_2 \, \bslash)}
\def\ialtb{{\tr}_{+}(\islash\, \aslash\, \lslash_2 \, \bslash)}
\def\bltloa{{\tr}_{+}(\bslash\, \lslash_2\, \lslash_1 \, \aslash)}
\def\jbloa{{\tr}_{+}(\jslash\, \bslash\, \lslash_1 \, \aslash)}
\def\ibPb{{\tr}_{+}(\islash\, \bslash\, \Pslash \, \bslash)}
\def\ijltb{{\tr}_{+}(\islash\, \jslash\, \lslash_2 \, \bslash)}

\def\ijloa{{\tr}_{+}(\islash\, \jslash\,  \lslash_1 \, \aslash)}
\def\ijblt{{\tr}_{+}(\islash\, \jslash\,  \bslash \, \lslash_2)}

\def\jakb{{\tr}_{+}(\jslash\, \aslash\, \kslash \, \bslash)}
\def\iakb{{\tr}_{+}(\islash\, \aslash\, \kslash \, \bslash)}

\def\tofo{{\tr}_{+}(\onslash\, \thslash\, \twslash \, \foslash)}
\def\foto{{\tr}_{+}(\onslash\, \thslash\, \foslash \, \twslash)}
\def\tofi{{\tr}_{+}(\onslash\, \thslash\, \twslash \, \fislash)}
\def\fito{{\tr}_{+}(\onslash\, \thslash\, \fislash \, \twslash)}

\def\lrangle#1#2{\langle #1\,#2\rangle}

\def\Li{{$\rm Li}_2$}
\def\eps{\epsilon}
\def\epsuv{{\epsilon_{\rm \mbox{\tiny UV}}}}
\let\bm=\bibitem
\let\la=\label

\def\npb#1#2#3{Nucl. Phys. {\bf{B#1}} #3 (#2)}
\def\plb#1#2#3{Phys. Lett. {\bf{#1B}} #3 (#2)}
\def\prl#1#2#3{Phys. Rev. Lett. {\bf{#1}} #3 (#2)}
\def\prd#1#2#3{Phys. Rev. {D \bf{#1}} #3 (#2)}
\def\cmp#1#2#3{Comm. Math. Phys. {\bf{#1}} #3 (#2)}
\def\cqg#1#2#3{Class. Quantum Grav. {\bf{#1}} #3 (#2)}
\def\nppsa#1#2#3{Nucl. Phys. B (Proc. Suppl.) {\bf{#1A}}#3 (#2)}
\def\ap#1#2#3{Ann. of Phys. {\bf{#1}} #3 (#2)}
\def\ijmp#1#2#3{Int. J. Mod. Phys. {\bf{A#1}} #3 (#2)}
\def\rmp#1#2#3{Rev. Mod. Phys. {\bf{#1}} #3 (#2)}
\def\mpla#1#2#3{Mod. Phys. Lett. {\bf A#1} #3 (#2)}
\def\jhep#1#2#3{J. High Energy Phys. {\bf #1} #3 (#2)}
\def\atmp#1#2#3{Adv. Theor. Math. Phys. {\bf #1} #3 (#2)}
\newcommand{\EQ}[1]{\begin{equation} #1 \end{equation}}
\newcommand{\AL}[1]{\begin{subequations}\begin{align} #1 \end{align}\end{subequations}}
\newcommand{\SP}[1]{\begin{equation}\begin{split} #1 \end{split}\end{equation}}
\newcommand{\ALAT}[2]{\begin{subequations}\begin{alignat}{#1} #2 \end{alignat}
                        \end{subequations}}
\def\beqa{\begin{eqnarray}}
\def\eeqa{\end{eqnarray}}
\def\beq{\begin{equation}}
\def\eeq{\end{equation}}
\def\sst{\scriptscriptstyle}
\def\thetabar{\bar\theta}
\def\Tr{{\rm Tr}}
\def\one{\mbox{1 \kern-.59em {\rm l}}}
 \def\Nh{\hat{N}}

\newcommand{\half}{{\textstyle {1 \over 2}}}

\def\a{\alpha}      \def\da{{\dot\alpha}}
\def\b{\beta}       \def\db{{\dot\beta}}
\def\c{\gamma}  \def\G{\Gamma}  \def\cdt{\dot\gamma}
\def\d{\delta}  \def\D{\Delta}  \def\ddt{\dot\delta}
\def\e{\epsilon}        \def\vare{\varepsilon}
\def\f{\phi}    \def\F{\Phi}    \def\vvf{\f}
\def\h{\eta}
\def\k{\kappa}
\def\l{\lambda} \def\L{\Lambda}
\def\m{\mu} \def\n{\nu}
\def\o{\omega}
\def\p{\pi} \def\P{\Pi}
\def\r{\rho}
\def\s{\sigma}  \def\S{\Sigma}
\def\t{\tau}
\def\th{\theta} \def\Th{\Theta} \def\vth{\vartheta}
\def\X{\Xeta}
\def\z{\zeta}
\def\de{\partial}

\def\cA{{\cal A}} \def\cB{{\cal B}} \def\cC{{\cal C}}
\def\cD{{\cal D}} \def\cE{{\cal E}} \def\cF{{\cal F}}
\def\cG{{\cal G}} \def\cH{{\cal H}} \def\cI{{\cal I}}
\def\cJ{{\cal J}} \def\cK{{\cal K}} \def\cL{{\cal L}}
\def\cM{{\cal M}} \def\cN{{\cal N}} \def\cO{{\cal O}}
\def\cP{{\cal P}} \def\cQ{{\cal Q}} \def\cR{{\cal R}}
\def\cS{{\cal S}} \def\cT{{\cal T}} \def\cU{{\cal U}}
\def\cV{{\cal V}} \def\cW{{\cal W}} \def\cX{{\cal X}}
\def\cY{{\cal Y}} \def\cZ{{\cal Z}}

\def\ua{\underline{\alpha}}
\def\ub{\underline{\phantom{\alpha}}\!\!\!\beta}
\def\uc{\underline{\phantom{\alpha}}\!\!\!\gamma}
\def\um{\underline{\mu}}
\def\ud{\underline\delta}
\def\ue{\underline\epsilon}
\def\una{\underline a}\def\unA{\underline A}
\def\unb{\underline b}\def\unB{\underline B}
\def\unc{\underline c}\def\unC{\underline C}
\def\und{\underline d}\def\unD{\underline D}
\def\une{\underline e}\def\unE{\underline E}
\def\unf{\underline{\phantom{e}}\!\!\!\! f}\def\unF{\underline F}
\def\unm{\underline m}\def\unM{\underline M}
\def\unn{\underline n}\def\unN{\underline N}
\def\unp{\underline{\phantom{a}}\!\!\! p}\def\unP{\underline P}
\def\unq{\underline{\phantom{a}}\!\!\! q}
\def\unQ{\underline{\phantom{A}}\!\!\!\! Q}
\def\unH{\underline{H}}

\def\As {{A \hspace{-6.4pt} \slash}\;}
\def\bs {{b \hspace{-6.4pt} \slash}\;}
\def\Ds {{D \hspace{-6.4pt} \slash}\;}
\def\ds {{\del \hspace{-6.4pt} \slash}\;}
\def\ss {{\s \hspace{-6.4pt} \slash}\;}
\def\ks {{ k \hspace{-6.4pt} \slash}\;}
\def\ps {{p \hspace{-6.4pt} \slash}\;}
\def\pas {{{p_1} \hspace{-6.4pt} \slash}\;}
\def\pbs {{{p_2} \hspace{-6.4pt} \slash}\;}
\def\Ps {{P \hspace{-6.4pt} \slash}\;}
\def\Qs {{Q \hspace{-6.4pt} \slash}\;}

\def\Fh{\hat{F}}
\def\Vh{\hat{V}}
\def\Xh{\hat{X}}
\def\ah{\hat{a}}
\def\xh{\hat{x}}
\def\yh{\hat{y}}
\def\ph{\hat{p}}
\def\xih{\hat{\xi}}
\def\psit{\tilde{\psi}}
\def\Psit{\tilde{\Psi}}
\def\tht{\tilde{\th}}
\def\lt{\tilde{\lambda}}
\def\hl{\hat{\lambda}}
\def\hlt{\hat{\tilde{\lambda}}}
\def\llt{\tilde{l}}
\def\At{\tilde{A}}
\def\Qt{\tilde{Q}}
\def\Rt{\tilde{R}}
\def\Nt{\tilde{N}}

\def\at{\tilde{a}}
\def\st{\tilde{s}}
\def\ft{\tilde{f}}
\def\pt{\tilde{p}}
\def\qt{\tilde{q}}
\def\vt{\tilde{v}}
\def\nt{\tilde{n}}

\def\delb{\bar{\partial}}
\def\bz{\bar{z}}
\def\bD{\bar{D}}
\def\bB{\bar{B}}

\def\bk{{\bf k}}
\def\bl{{\bf l}}
\def\bp{{\bf p}}
\def\bq{{\bf q}}
\def\br{{\bf r}}
\def\bx{{\bf x}}
\def\by{{\bf y}}
\def\bR{{\bf R}}
\def\bV{{\bf V}}

\def\d{\delta}\def\D{\Delta}\def\ddt{\dot\delta}
\def\pa{\partial} \def\del{\partial}
\def\xx{\times}
\def\uno{\mbox{1 \kern-.59em {\rm l}}}
\def\trp{^{\top}}
\def\inv{^{-1}}
\def\dag{{^{\dagger}}}
\def\pr{^{\prime}}
\def\lan{\langle}
\def\ran{\rangle}
\def\rar{\rightarrow}
\def\lar{\leftarrow}
\def\lrar{\leftrightarrow}
\newcommand{\0}{\,\!}      
\def\one{1\!\!1\,\,}
\def\im{\imath}
\def\jm{\jmath}
\newcommand{\tr}{\mbox{tr}}
\newcommand{\slsh}[1]{/ \!\!\!\! #1}
\def\vac{|0\rangle}
\def\lvac{\langle 0|}
\def\hlf{\frac{1}{2}}
\def\ove#1{\frac{1}{#1}}
\def\Box{\square}
\def\ZZ{\mathbb{Z}}
\def\CC#1{({\bf #1})}
\def\bcomment#1{}
\def\bfhat#1{{\bf \hat{#1}}}
\def\VEV#1{\left\langle #1\right\rangle}
\newcommand{\ex}[1]{{\rm e}^{#1}} \def\ii{{\rm i}}
\def\rr{{\rm r}} \def\rs{{\rm s}}\def\rv{{\rm v}}
\def\ri{{\rm i}}\def\rj{{\rm j}}
\newcommand{\lrbrk}[1]{\left(#1\right)}
\newcommand{\sfrac}[2]{{\textstyle\frac{#1}{#2}}}

\def\Li{{\rm Li}_2}

\font\mybb=msbm10 at 12pt
\def\bb#1{\hbox{\mybb#1}}

\font\myBB=msbm10 at 18pt
\def\BB#1{\hbox{\myBB#1}}

\begin{document}

\begin{flushright}
QMUL-PH-09-06
\end{flushright}

\vspace{20pt}

\begin{center}

{\Large \bf One-Loop Amplitudes in $\mathcal{N}=4$ Super Yang-Mills   }
\\
\vspace{0.3cm}
{\Large \bf  and  Anomalous Dual Conformal Symmetry    }
\vspace{11pt}
\vspace{32pt}

{\mbox {\bf Andreas Brandhuber, Paul Heslop and Gabriele Travaglini}}%
\footnote{
{\sffamily \{\tt a.brandhuber, p.j.heslop, g.travaglini\}@qmul.ac.uk }}

{\em Centre for Research in String Theory\\
Department of Physics\\
Queen Mary, University of London\\
Mile End Road, London, E1 4NS\\
United Kingdom
 }

\vspace{30pt} {\bf Abstract}

\end{center}

\noindent
We discuss what predictions can be made for one-loop superamplitudes in maximally supersymmetric Yang-Mills theory by using anomalous dual conformal symmetry.  We show that the anomaly coefficient is a specific combination of two-mass hard 
and one-mass supercoefficients which appears in the supersymmetric on-shell recursion relations and equals the corresponding tree-level superamplitude.  We discuss further novel relations among supercoefficients imposed by the remaining 
non-anomalous part of the symmetry. 
In particular, we find that all one-loop supercoefficients, except the four-mass box coefficients, can be expressed as linear combinations of three-mass box coefficients and a particular symmetric combination of two-mass hard coefficients.
We check that our equations are explicitly satisfied in the case of one-loop $n$-point MHV and NMHV amplitudes. 
As a bonus, we prove the covariance of the NMHV superamplitudes at an arbitrary number of points, 
extending previous results at  $n\leq 9$.

\noindent

\setcounter{page}{0}
\thispagestyle{empty}
\newpage


\section{Introduction  }
\setcounter{footnote}{0}

Recently, a new symmetry of planar scattering amplitudes in $\cN=4$ super Yang-Mills (SYM) has been proposed  
in \cite{dhks},  called dual superconformal symmetry. 
This symmetry is expected to be exact at tree level, and violated by an anomaly at the quantum  level. 
Indeed, it was proved  in \cite{bhtrec}  that  the tree-level  $S$-matrix of the planar $\cN=4$ theory is covariant under 
the dual superconformal symmetry, and similar covariance properties were also established for the supercoefficients 
of the one-loop expansion of the scattering amplitudes in a basis of box functions \cite{bhtrec,dhksgen}. 

On the string theory side, the origin of this symmetry has been explained neatly in 
\cite{bermal,tse}  using a T-duality of the  superstring theory on $AdS_5 \times S^5$, 
which involves a bosonic T-duality \cite{am}   and fermionic T-duality. 
A certain combination of these bosonic and fermionic T-dualities maps 
 the original string sigma model into a dual sigma model 
which turns out to be identical to the original one. Moreover, the T-duality exchanges the original with the dual superconformal symmetries. 

In contrast to the ordinary superconformal symmetry of the maximally supersymmetric theory, its dual counterpart is not a symmetry of the action but only of the amplitudes at the planar level.  Interestingly, the authors of  \cite{dhp} considered the commutation relations between ordinary and dual superconformal generators and discovered that these give rise to a 
Yangian symmetry at tree level. Constraints imposed on amplitudes by the ordinary superconformal symmetry have recently been considered in \cite{new}.

An important feature of the strong coupling calculation of scattering amplitudes in the dual sigma model 
is that it is identical to that of a Wilson loop with a special polygonal contour,  
constructed by gluing together the  lightlike momenta of the scattered particles  
following the order of the insertions of the string vertex operators \cite{am}. 
Strikingly, there is now growing evidence that  calculations in  weakly coupled $\cN=4$ SYM  
of the same Wilson loops -- specifically at  one \cite{dks,bht} and two loops \cite{dhks4,dhks5,dhksbum,dhks6} --  
are in perfect  agreement  with the MHV scattering amplitudes of the $\mathcal{N}=4$ theory
calculated in \cite{bddk,abdk,2l5pt,seven}.  A numerical calculation of Wilson loops  at two loops for an 
arbitrary number of edges, or scattered particles, has been carried out in \cite{Anastasiou:2009kn}, 
and awaits explicit results for the corresponding amplitudes in $\cN=4$ SYM. 

At weak coupling, the origin of the dual conformal symmetry was  understood  from the Wilson loop perspective in \cite{dhks5}, where it was noted that it is nothing but the ordinary conformal symmetry of the Wilson loop acting on the 
't Hooft's region (or T-dual) momenta  $x_i$'s, defined via 
\beq
p_{i, \a \dot{\a}} \ = \ (x_{i}- x_{i+1})_{\a \dot{\a}}
\ , 
\eeq
where $i=1, \ldots , n$. Here $n$ is the number of scattered particles,  and we identify $x_{n+1} = x_1$. 
Importantly, due to the presence of cusps in the polygonal contour, the symmetry is anomalous. 
In \cite{dhks4}, an expression for the  Wilson loop anomaly  was proposed at one loop, 
and later extended and proved to all loops in \cite{dhks5}.
Furthermore, the  four- and five-point Wilson loops
are completely  determined (up to a constant) by  the anomalous dual conformal Ward identity \cite{dhks5}, 
and found to be of the form predicted by the ABDK/BDS ansatz for the amplitudes \cite{abdk,bds}.  

Based on the connection between Wilson loops and MHV amplitudes,  
an all-loop expression of the anomaly of generic loop amplitudes was proposed in \cite{dhks}. 
Schematically, this anomaly is proportional to the corresponding tree-level amplitude, multiplied by the all-loop anomaly
of the Wilson loop.

In this paper we wish to investigate the origin of this anomaly on the amplitude side, and use the expression of the anomaly conjectured in \cite{dhks}  to make predictions for the supercoefficients of 
generic (in particular also  non-MHV)  one-loop superamplitudes. 
Specifically, our strategy will consist in applying the dual conformal generators to a generic superamplitude, which can be written \cite{bddk}  as a linear combination of box functions times supercoefficients. We will then use the covariance of the one-loop supercoefficients proved in \cite{bhtrec} and \cite{dhks}, which allows 
the conformal generators to pass through the supercoefficients and thus act on the box functions.  

Now, box functions are  (with the exception of the four-mass box) infrared divergent,  and need to be regularised, which is usually accomplished by  evaluating them in $4-2\epsilon$ dimensions. 
However, if one could work in four dimensions (by e.g.~resorting to a ``regularisation" of the box which makes  the external kinematics massive), the boxes would be invariant under the dual conformal symmetry  \cite{magic}  -- a fact that has been referred to as ``pseudo-conformality" of these integral functions, and has played an important role in higher-loop calculations  of 
$\cN=4$ MHV amplitudes
\cite{Cachazo:2006tj,2l5pt,Bern:2006ew,Cachazo:2006az,Bern:2007ct,Spradlin:2008uu,Vergu:2009zm}.
Dimensional regularisation cleanly exposes the anomalies of  the box functions. 
We will therefore calculate all these box  anomalies, and use them in order to write down the general expression for the anomaly of an arbitrary  non-MHV amplitude. This turns out to  be proportional to a particular linear combination of supercoefficients. 

The equations we derive exhibit two important features. Firstly,  the one-loop anomaly conjectured in \cite{dhks} 
arises naturally in our setup as the particular combination of two-mass hard and one-mass coefficients considered in \cite{dissolving}, which is equal to the tree-level amplitude.  
Furthermore, we find a new set of relations which need to be satisfied if the expression of the all-loop anomaly of \cite{dhks} is correct. By assuming that this is indeed the case, we derive new equations which we use  to relate supercoefficients of  
one-loop superamplitudes. Specifically,  we find that we can re-express all supercoefficients (except the four-mass ones) in terms of three-mass box coefficients and a particular symmetric combination of two-mass hard coefficients. Four-mass coefficients multiply integrals which are finite \cite{mand} and dual conformal invariant  \cite{magic}, hence we are not  able to constrain them. 

Finally, we will present some checks of the new equations we propose.  Specifically, we will consider the infinite sequences of MHV and NMHV superamplitudes in $\cN=4$ SYM computed in \cite{bddk} and \cite{dhksgen}, respectively, 
and prove that  their expressions satisfy our conformal equations. 
The proof of dual conformal invariance in the NMHV case requires for $n\geq10$ particles a new identity for the superconformal invariant $R_{rst} $ in terms of which the NMHV superamplitude was expressed in \cite{dhksgen}. 
We show this in an Appendix to this paper.  A further brief Appendix is devoted to constructing combinations of box functions which are dual conformal invariant.

The rest of the paper is organised as follows.%
\footnote{The results of this paper were announced at the ``International Workshop on Gauge and String Amplitudes"
at Durham University, 30 March--3 April 2009 \cite{slides}.}
In the next Section, after writing the expansion of generic amplitudes in terms of box functions to review our notation, we will calculate the anomalies of all types of one-loop box functions. 
We will explain how this anomaly arises from  an anomalous conformal Jacobian, which formally vanishes as $\eps \to 0$, but leads to a nonvanishing contribution because of the presence of infrared divergences.
We will use these results to calculate, in Section 3, the expression of the dual conformal anomaly of an arbitrary $n$-point one-loop superamplitude in terms of the supercoefficients. 
In Section 4 we analyse the conformal equations. Firstly, we show that these equations include, as a subset, the infrared consistency conditions, and we present a way to disentangle the new equations from the infrared equations. Then, we show how these equations can in principle be solved in terms of 
three-mass box coefficients and a  symmetric combination of two-mass hard coefficients.
Finally, 
we check in Section 5 that our equations are explicitly satisfied in the case of one-loop $n$-point MHV and NMHV amplitudes.

{\bf Note added:}
After finishing this work, we were made aware in recent email correspondence with 
Henriette Elvang and Dan Freedman of a work of Elvang, Freedman and Kiermaier \cite{EFK} 
which has some overlap with our paper.

\section{One-loop amplitudes and anomalous box functions}
In this section we start off by introducing the expansion of generic $n$-point superamplitudes in terms of box functions. We will then move on to derive the dual conformal anomalies of the different box functions, i.e.~zero-mass, one-mass, two-mass easy, two-mass hard, and three-mass. The four-mass box is conformal invariant, and we will not have anything to say about the supercoefficients of these functions.

\subsection{One-loop superamplitudes}

In the supersymmetric formalism of \cite{Nair},  to each particle in the $\cN=4$ theory one associates 
commuting spinors
$\l_\a$, $\lt_{\da}$ (in terms of which the momentum of the $i^\mathrm{th}$
particle is $p_{\a \da}^{i} = \l_{\a}^{i} \lt_{\da}^{i}$), as well
as anticommuting variables $\eta^{A}_{i}$, where $A=1,\ldots,4$ is an $SU(4)$ index. 
The supersymmetric amplitude can then be expanded in powers of the
$\cN=4$ superspace coordinates $\eta_{A}^{i}$ for the different particles, 
and each term of this expansion corresponds to a particular scattering
amplitude in $\cN=4$ SYM with a fixed total helicity $h_\mathrm{tot}=\sum_{i=1}^n h_i $. A term containing $m_i$ powers of
$\eta_{A}^{i}$ corresponds to a scattering process where the
$i^\mathrm{th}$  particle has helicity $h_i = 1 - m_i/2$. 
Explicitly, the $n$-point  MHV superamplitude is \cite{Nair} 
\beq
\label{NairMHV}
\cA_{\rm MHV} (1, \ldots , n) \ = \ i (2\pi)^4 \, 
{\delta^{(4)} (\sum_{i=1}^{n} \l_i \lt_i ) \, \delta^{(8)} ( \sum_{i=1}^{n} \eta_i \l_i ) 
\over \lan12\ran \cdots \lan n1\ran } 
\ , 
\eeq
where, as usual, $\lan i  j\ran := \eps_{\a \b} \l_i^\a \l_j^\b$, 
and the two delta functions in the numerator impose
momentum and supermomentum conservation, respectively. 
The use of superamplitudes is not only conceptually important,  
but also has  the practical advantage of  allowing for efficient ways to perform sums over internal helicities which occur at tree level and in (generalised) unitarity cuts   
\cite{Georgiou:2004wu,bst, Massimo,Elvang:2008na,Elvang:2008vz,dhksgen,cahk,Bern:2009xq}.

\begin{figure}[ht]
\begin{center}
\scalebox{0.55}{\includegraphics{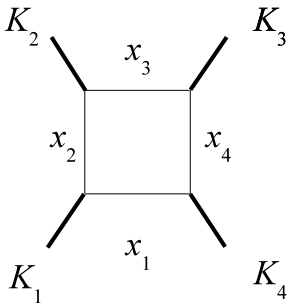}}
\end{center}
\caption{\it  A generic box function. 
$K_1, K_2, K_3$ and $K_4$  denote the  external momenta, and $x_1, x_2, x_3$ and $x_4$  the
corresponding region momenta, with $K_i = x_i - x_{i+1}$, $i=1, \ldots , 4$. }
\label{boxfunction}
\end{figure}

Let us now describe the structure of one-loop superamplitudes.
The natural basis to consider in the context of dual conformal symmetry 
is  given by the scalar box functions $F_i$.
These  are related to the 
scalar box integrals $I_i$ by a kinematic prefactor \cite{bddk}, in the following way. 
We call  $K_1, K_2, K_3$ and $K_4$  the external momenta at the four corners of a given box function,  
which are expressed as sums of momenta $p_i$ of external
particles. The momenta 
$K_{1\ldots4}$ can also be written  in terms of the region momenta 
$x_{1\ldots4}$, e.g. $K_1 = x_{12}$, where $x_{ij} := x_i-x_j$ (see Figure \ref{boxfunction}).
Then, up to a numerical constant, the relation between the $F$'s and the $I$'s  is%
\footnote{In \eqref{bpm}  we use a collective index $i$ to denote the box function with external momenta 
$K_{1\ldots 4}$.}
\begin{eqnarray}
\label{bpm}
I_i &= &-2\, \frac{F_i}{\sqrt{R_i}} \ , \nonumber \\
R_i&=& (x_{13}^2 x_{24}^2)^2 - 2 x_{13}^2 x_{24}^2  x_{12}^2  x_{34}^2  
- 2 x_{13}^2 x_{24}^2  x_{23}^2  x_{41}^2 +(x_{12}^2 x_{34}^2 - x_{23}^2 x_{41}^2)^2 \ .
\end{eqnarray}
In this paper we will not make any statement on coefficients of four-mass box functions, 
as these are infrared finite and trivially invariant under dual conformal symmetry. 
We can then  simplify the expression for $\sqrt{R}$ to 
\beq 
\sqrt{R} \ \to \ x_{13}^2 x_{24}^2 - x_{23}^2 x_{41}^2 
\ , 
\eeq
valid for all box functions except four-mass ones in the case 
where either $x_{12}^2$ or $x_{34}^2$ vanish. Notice that,  under dual conformal  inversions, one has 
\be\label{rtrafo}
\sqrt{R_i} \ \to \  \frac{\sqrt{R_i}}{x_1^2 x_2^2 x_3^2 x_4^2} \, .
\ee
We  expand   a generic  $n $-point one-loop superamplitude $\mathcal{A}^{\mathrm{1-loop}}_n $ in terms of box functions \cite{bddk} as 
\begin{equation}
\label{exp3}
 {\mathcal{A}}^{\mathrm{1-loop}}_n \ =  \  
\sum_{\{i , j , k , l\}}  
{c}(i,j,k,l)  \,
F(i,j,k,l) \ ,
\end{equation}
where $i$, $j$, $k$, $l$, denote the four region momenta of the box function (as in Figure \ref{boxfunction}, with 
the labels $1$, $2$, $3$, $4$, replaced by $i$, $j$, $k$, $l$).

In \cite{bhtrec} it was shown that the supercoefficients 
$c_i$  have uniform covariant transformation properties under
dual conformal transformations just as the corresponding tree-level amplitudes. 
Specifying to the different types of box functions, we get 
\beqa
\label{exp2}
\mathcal{A}^{\mathrm{1-loop}}_n  & = &    
\sum_i  c^\mathrm{1m}(i)  \, F^\mathrm{1m}_i \,  + \, 
\sum_{i,j}  c^\mathrm{2me}(i,j)  \, F^\mathrm{2me}_{ij} \,  \nonumber \\
&+& \sum_{i,j}  c^\mathrm{2mh}(i,j)  \, F^\mathrm{2mh}_{ij} \,  + \, 
\sum_{i,j,k}  c^\mathrm{3m}(i,j,k)  \, F^\mathrm{3m}_{ijk} \,  + \, 
\cdots \ , 
\eeqa
where the dots stand for the four-mass box contributions. 
The sums in \eqref{exp2}  are extended over the $n$ one-mass, $n(n-5)/2$ two-mass easy, 
$n(n-5)$ two-mass hard, and $n(n-5)(n-6)/2$ three-mass box functions. Furthermore, the
coefficients of the various types of box functions are related to the general
box coefficients as 
\beqa
&&c^\mathrm{1m}(i) = c(i,i+1,i+2,i+3) \ , \quad 
c^\mathrm{2me}(i,j) = c(i,i+1,j,j+1) \ , \ \nonumber \\
&&c^\mathrm{2mh}(i,j) = c(i,i+1,i+2,j) \ , \quad
c^\mathrm{3m}(i,j,k) = c(i,i+1,j,k) \ .
\eeqa
In  \cite{Bern:2004bt} it was observed that the total number of 2me and 1m coefficients, $n(n-3)/2$, precisely equals the number 
of infrared consistency equations. Furthermore, it was also noticed that for $n$ odd these equations are independent and can therefore 
be used to determine all the 2me and 1m coefficients in terms of the 2mh and 3m coefficients; whereas, for $n$ even, it was checked up to $n=29$ that one equation is actually redundant.

\subsection{Anomalous one-loop amplitudes}
Tree-level superamplitudes are covariant under dual superconformal symmetry, as  conjectured in 
\cite{dhks} and later proved in \cite{bhtrec}. 
In order to deal with quantities which are invariant under dual conformal transformations (rather than covariant), 
we follow \cite{dhp} and redefine the dual conformal generator $K^{\mu}$ to be 
\beq 
\label{khat}
K^\mu \ \to \ \hat{K}^\mu \ := \ K^{\mu} + \sum_{i=1}^n x^\mu_i
\ . 
\eeq
Then $\hat{K}^\mu$ annihilates any tree-level superamplitude.

At one loop,  amplitudes are no longer  covariant under
four-dimensional conformal transformations as they are at tree level because 
of the presence of infrared  divergences. 
Instead, according to the  conjecture of \cite{dhks}, 
they are expected to  have the following anomaly at one loop under a dual conformal transformation:
\begin{align}\label{eq:2}
\hat{K}^\mu    {\mathcal{A}}^{\mathrm{1-loop}}_n = 4\epsilon\,
 {\mathcal{A}}^{\mathrm{tree}}_n \sum_{i=1}^n x_{i+1}^\mu x_{i\, i+2}^2 \,J(x_{i\, i+2}^2)\ , 
\end{align}
where the one-mass triangle $J(a)$ is defined by 
\beq
J(a) \:= \ {r_\Gamma\over \epsilon^2 }
 (-a)^{-\epsilon -1} \ ,  
\eeq
and $r_\Gamma := \Gamma (1 + \e) \Gamma^2 ( 1 - \e )  / \Gamma (1 - 2 \e) $.


We then apply $\hat{K}^\mu$  defined  in \eqref{khat} to both sides of \eqref{exp3} to determine the
consequences of dual conformal symmetry. 
Acting upon the left hand side
we get the dual conformal anomaly in  \eqref{eq:2}. 
Acting upon the right hand side,   we get 
$  
\sum_{\{i , j , k , l\}}  
{c}(i,j,k,l)  \,K^\mu
F(i,j,k,l) $ 
since $\hat{K} c=0$,   which is the statement that  all  box coefficients are covariant  under dual conformal symmetry, hence
\beq
\label{2.9}
\hat{K}^\mu  {\mathcal{A}}^{\mathrm{1-loop}}  \ = \ 
\sum_{\{i , j , k , l\}}  
{c}(i,j,k,l)  \  K^\mu
F(i,j,k,l) \ . 
\eeq
It is therefore clear that the anomaly stems entirely from the (anomalous) box functions. Our next task will then consist in calculating explicitly the dual conformal anomalies of all the one-loop box functions.

A comment is in order here. 
The statement of the anomalous conformal invariance \eqref{eq:2} is not
phrased in the same way as that of  \cite{dhks} but the two
statements can easily
be shown to be equivalent. 
In \cite{dhks},  dual superconformal invariance 
was used to argue that  any superamplitude can be written as the MHV superamplitude, which captures the anomaly, 
times a dual superconformal  invariant factor $\cR$, i.e.
\begin{align}
   \cA =  {\cA}_\mathrm{MHV} \,  \cR \ .
\end{align}
Expanding this out in the coupling constant
 we obtain
\begin{align}
   {\cA}^\mathrm{tree} &=\  {\cA}^\mathrm{tree}_\mathrm{MHV}\cR^\mathrm{tree} \ , 
   \\
   {\cA}^\mathrm{1-loop} &= \ {\cA}^\mathrm{tree}_\mathrm{MHV}\cR^\mathrm{1-loop}\,  +\,    
  {\cA}^\mathrm{1-loop}_\mathrm{MHV} \cR^\mathrm{tree}\ .\label{eq:3}
\end{align}
The first equation is simply the statement that all tree-level amplitudes
 are covariant under dual
conformal symmetry which was proved in \cite{bhtrec}.
As for the second equation, by applying a conformal  transformation to both sides
of \eqref{eq:3}  and
using the  anomalous transformation of the one-loop MHV
amplitude \cite{dhks4}, which follows from the Wilson loop/amplitude
duality, 
\begin{align}
  \hat{K}^\mu  {\cA}^\mathrm{1-loop}_\mathrm{MHV}  &= 4 \epsilon\, {\cA}^\mathrm{tree}_\mathrm{MHV}\, \sum_{i=1}^n x_{i+1}^\mu x_{i\,
  i+2}^2 \,J(x_{i\, i+2}^2)\ ,
\end{align}
we obtain directly \eqref{eq:2}.

\subsection{Anomalies of  one-loop box functions}

Here we derive the anomaly of generic box functions under dual conformal transformations, which we will then use to study 
\eqref{2.9}. As mentioned earlier, four-mass box functions are invariant under dual conformal symmetry \cite{magic}, 
hence we will not need to consider them. 

Consider a generic  one-loop box function, 
\beq
F \ = \ -i{ (4 \pi)^{2- \epsilon}}
\int\!\!{d^Dx_5 \over (2\pi)^D}\, {(-\sqrt{R}/ 2) \over x_{51}^2 x_{52}^2 x_{53}^2 x_{54}^2} \ , 
\eeq
where $\sqrt{R}$ is defined in \eqref{rtrafo} and $D = 4 - 2 \epsilon$.  
We perform a dual conformal transformation on the region momenta of the box function, 
\beq
\label{specialconf}
x_i^\mu \ \to  x_i^{\prime \, \mu}\ := \ \ {x_i^\mu + b^\mu x_i^2 \over 1 + 2 (b x_i) + x_i^2 b^2} \ , 
\eeq
where $i =1, \ldots, 4$. 
This transformation can be compensated by performing an identical  one on the integration variable $x_5$. 
Doing so, one obtains%
\footnote{Recall that  differences of dual momenta transform covariantly, $
(x-y)^2 \to  (x-y)^2  / \big[ \big(1 + 2 (bx) + 2 b^2 x^2\big) \big(1 + 2 (by) + 2 b^2 y^2\big)\big] $.}
\beq 
F( x_1^\prime, x_2^\prime, x_3^\prime, x_4^\prime ) \ = \ 
 -i{ (4 \pi)^{2- \epsilon}}
\int\!\!{d^Dx_5 \over (2\pi)^D}\,   \, |J| \, {(-\sqrt{R}/2)  \over  x_{51}^2 x_{52}^2 x_{53}^2 x_{54}^2} 
\left( 1 + 2 (bx_5) + b^2 x_5 ^2\right)^4  
\ , 
\eeq
where 
\beq 
|J| \, := \, \mathrm{det}_{\m, \n} \left( {\partial x_5^{\prime\, \mu} \over \partial x_5^\nu}\right)  \ = \ 1 \, - \, 2 D (bx_5) \, + \, \mathcal{O} (b^2)
\ .
\eeq
 $x_5^\prime$ is given by an expression identical to \eqref{specialconf}, and 
we have used $\eta_{\mu \nu} \eta^{\mu \nu} = D$. 
We conclude that, under a dual conformal transformation, the variation of 
a generic box function is 
\beq
\label{pvreduce}
\d_b F \ = \ 2 (4 - D) \big[   -i{ (4 \pi)^{2- \epsilon}} \big]
\int\!\!{d^Dx_5 \over (2\pi)^D}\,  \, {(-\sqrt{R}/2)\over x_{51}^2 x_{52}^2 x_{53}^2 x_{54}^2}  ( b x_5) \ .  
\eeq
The integral appearing on the right hand side of \eqref{pvreduce} is a linear box, which can 
be evaluated straightforwardly with a Passarino-Veltman (PV) reduction \cite{Passarino:1978jh}. 
Crucially we have $4-D = 2 \epsilon$, hence it will be  enough to pick the poles in inverse powers of $\epsilon$ 
in  the PV reduction appearing on the right hand side of \eqref{pvreduce}, which greatly facilitates our task.

 \begin{figure}[ht]
\begin{center}
\scalebox{0.57}{\includegraphics{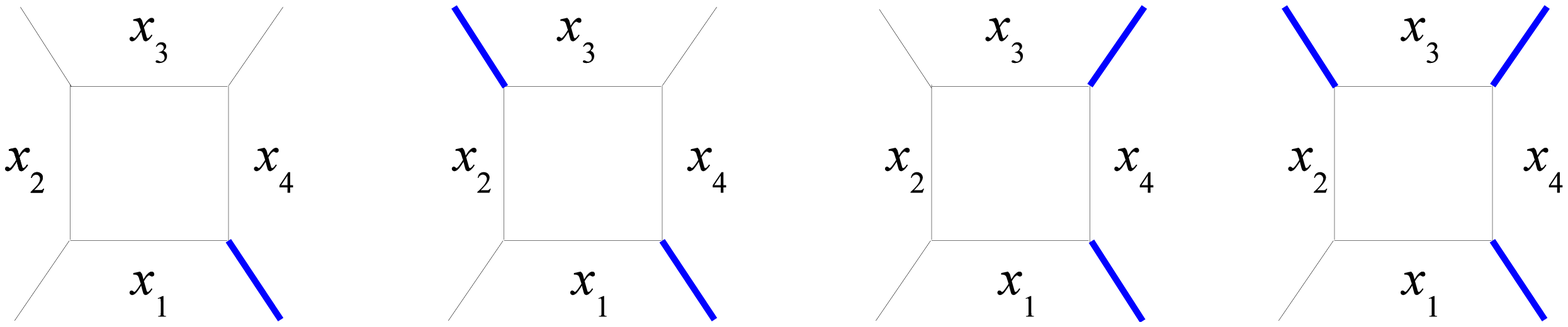}}
\end{center}
\caption{\it  The 1m, 2me, 2mh, 3m box functions whose dual conformal anomalies are calculated in 
\eqref{cieq}.  Massive legs are depicted in bold blue. }
\label{boxanomalies}
\end{figure}

We have performed the relevant PV reductions for the linear boxes appearing in \eqref{pvreduce}, 
which in turn give the anomalous action  of dual conformal generators on each type of box functions. 
Our results for the boxes represented in Figure \ref{boxanomalies} are the following: 
\begin{eqnarray}
 \label{cieq}
K^\mu F^\mathrm{0m} \!\!\!&=&\!\!\! 
2\epsilon \left[  (x_1 + x_3)^\mu x_{24}^2 J( x_{24}^2) \, + \, (x_2 + x_4)^\mu x_{13}^2 J (x_{13}^2) 
\right]
\ , 
\\ \nonumber
K^\mu F^\mathrm{1m}_1  \!\!\!&=&\!\!\! -2\epsilon \left\{ 
- x_1^\mu x_{24}^2 J(x_{24}^2 , x_{41}^2)  + x_2^\mu \Big[ x_{41}^2 J (x_{24}^2, x_{41}^2 ) - x_{13}^2 J (x_{13}^2) \Big] \right. 
\\ \nonumber 
 \!\!\!&+&\!\!\!  \left.x_3^\mu \Big[ x_{41}^2 J (x_{13}^2, x_{41}^2 ) - x_{24}^2 J (x_{24}^2) \Big]  
- x_4^\mu x_{13}^2 J(x_{13}^2 , x_{41}^2)\right\}\ , 
\\ \nonumber 
K^\mu F^\mathrm{2me}_{13}   \!\!\!&=&\!\!\!-2 \epsilon \left\{ x_1^\mu \Big[ - x_{24}^2 J(x_{24}^2 , x_{41}^2) + x_{23}^2 J(x_{13}^2 , x_{23}^2)\Big] 
+ x_2^\mu \Big[ - x_{13}^2 J(x_{13}^2 , x_{23}^2) + x_{41}^2 J(x_{24}^2 , x_{41}^2)\Big] \right. 
\nonumber \\ 
 \!\!\!&+&\!\!\! \left. x_3^\mu \Big[ - x_{24}^2 J(x_{24}^2 , x_{23}^2)  + x_{41}^2 J(x_{13}^2 , x_{41}^2)\Big] + 
x_4^\mu \Big[ - x_{13}^2 J(x_{13}^2 , x_{41}^2) +  x_{23}^2 J(x_{24}^2 , x_{23}^2)\Big] 
\right\}\nonumber 
\ , 
\\ \nonumber 
K^\mu F^\mathrm{2mh}_{14}  \!\!\!&=&\!\!\!-2 \epsilon \left\{ 
-x_1^\mu x_{24}^2  J( x_{24}^2 , x_{41}^2 ) \right. \\ \nonumber 
 \!\!\!&+&\!\!\! 
x_2^\mu \Big[  x_{41}^2 J(x_{24}^2 , x_{41}^2) - x_{13}^2 J(x_{13}^2)  + x_{34}^2 J (x_{24}^2, x_{34}^2 ) \Big]
\\ \nonumber   \!\!\!&-&\!\!\! 
\left. x_3^\mu x_{24}^2 J( x_{24}^2 , x_{34}^2 )  \Big] \right\}
\ , 
\\ \nonumber
K^\mu F^\mathrm{3m}_{134}  \!\!\!&=&\!\!\! -2\epsilon \left\{ 
x_1^\mu \Big[ x_{23}^2 J (x_{23}^2 , x_{13}^2) - x_{24}^2 J(x_{24}^2, x_{41}^2) 
\Big] 
+ x_2^\mu \Big[ x_{41}^2 J (x_{41}^2 , x_{24}^2) - x_{13}^2 J(x_{13}^2, x_{23}^2) \Big]
\right\}
\ . 
\end{eqnarray}
 The  two-mass triangle $J(a,b)$ is  defined by 
\begin{eqnarray}
J(a, b) &:= & {r_\Gamma\over \epsilon^2 }
 { (-a)^{-\epsilon} - (-b)^{-\epsilon}\over (-a) - (-b)   } 
\  . 
\end{eqnarray}
Notice also that  $J(a, b) \to J(a)$ as $b \to 0$.

Below we also list the corresponding infrared divergent terms for the same integral box functions: 
\begin{eqnarray}\label{eq:1}
\left.F^\mathrm{0m}\right|_{\mathrm{IR}} 
&=&- { r_\Gamma\over \epsilon^2 } 
\Big[ (-x_{13}^2)^{-\epsilon} +  (-x_{24}^2)^{-\epsilon} \Big] 
\ , 
\\ \nonumber 
\left.F_1^\mathrm{1m}\right|_{\mathrm{IR}} 
&=&- { r_\Gamma\over \epsilon^2 } 
\Big[ (-x_{13}^2)^{-\epsilon} +  (-x_{24}^2)^{-\epsilon}  - (-x_{41}^2)^{-\epsilon} \Big] 
\ , 
\\ \nonumber 
\left.F_{13}^\mathrm{2me}\right|_{\mathrm{IR}} 
&=& -{r_\Gamma\over \epsilon^2 } 
\Big[ (-x_{13}^2)^{-\epsilon} +  (-x_{24}^2)^{-\epsilon} - (-x_{23}^2)^{-\epsilon} - (-x_{41}^2)^{-\epsilon} \Big] 
\ , 
\\ \nonumber 
\left.F_{14}^\mathrm{2mh}\right|_{\mathrm{IR}} 
&=& -{r_\Gamma\over \epsilon^2 } 
\Big[ {1\over 2} {(-x_{13}^2)^{-\epsilon}}  +  (-x_{24}^2)^{-\epsilon} -{1\over 2} {(-x_{34}^2)^{-\epsilon}}  -{1\over 2} {(-x_{41}^2)^{-\epsilon}} 
 \Big] 
 \ , 
 \\ \nonumber 
\left.F_{134}^\mathrm{3m}\right|_{\mathrm{IR}} 
&=& -{r_\Gamma\over \epsilon^2 } \, 
\Big[{1\over 2} (-x_{13}^2)^{-\epsilon} + {1\over 2} (-x_{24}^2)^{-\epsilon} -{1\over 2} (-x_{23}^2)^{-\epsilon} - {1\over 2}(-x_{41}^2)^{-\epsilon} \Big] 
\ . 
\end{eqnarray}
A comparison of the infrared divergent terms of the boxes \eqref{eq:1} with the
conformal variations  \eqref{cieq} of the same integral functions shows that the former  
quantities can be  obtained from the latter  by simply  replacing every occurrence 
of $\epsilon \, x_i^\mu x_{jk}^2 J(x_{jk}^2, x_{ik}^2)$ in \eqref{cieq}
with $-r_{\Gamma} (-x_{jk}^2)^{-\epsilon}/(4\epsilon^2)\equiv x_{jk}^2 J(x_{jk}^2)/4$.  
We will make use of this observation later on to show that relations of the box
coefficients  which arise
from considering the infrared divergences of the amplitude are all
implied by the anomalous dual conformal symmetry.

\section{Conformal constraints on scattering amplitudes}
In this section we examine in more detail the constraints imposed by
dual conformal invariance  
on the coefficients of the one-loop superamplitudes. 
Anticipating our story a little, in the following we will find  that: 

{\bf 1. } The $n(n-3)/2$ infrared consistency equations are always implied by dual conformal invariance, hence they do not 
provide new relations; 

{\bf 2. } The number of independent conformal constraints we find is equal to $n(n-4)$, regardless of $n$ being even or odd.

Thus, conformal equations imply $n(n-4) - n(n-3)/2 = n(n-5)/2$ new constraints on top of the infrared consistency relations.  
We will see how the  conformal equations allow to eliminate $n$ one-mass box function coefficients, the $n(n-5)/2$ two-mass easy box coefficients, and $n(n-5)/2 $ antisymmetric combinations, to be specified later, of the two-mass hard coefficients. 
As a result, one is left with all the three-mass coefficients and a symmetric combination of two-mass hard coefficients which are 
the real unknowns of the problems (and can be calculated using quadruple cuts).

\subsection{ Generic $n$-point superamplitudes}

We now move on to  derive the constraints on the box
coefficients from dual conformal invariance for  general $n$-point amplitudes. 
To illustrate our results in concrete examples, we will then  briefly consider 
the cases  $n= 4,5,6$.

We start our analysis from \eqref{2.9}, and we use the results \eqref{cieq}
 for the anomalies of the box functions
derived in \eqref{cieq}. 
Examining carefully the variation of the boxes
under dual conformal transformations \eqref{cieq},  we see that all terms
will be either of the form 
\beq
x_{i}^{\mu}\, x_{i-1\,k}^2\,J(x_{i k}^2,x_{i-1\, k}^2) \ , \ \ \mathrm{ or}  \ \ \  
x_{i-1}^{\mu}\, x_{ik}^2\,J(x_{ik}^2,x_{i-1\, k}^2)
\ , 
\eeq
for some $i,k$.
These terms appear in the variations of the box functions  $F(i,k,j,i-1)$ and
$F(i,j,k,i-1)$ only, for all allowed values of $j$. 
Examining further the sign with which these terms appear we find that  
\beqa
\hat{K}^{\mu}  {\mathcal{A}}^{\mathrm{1-loop}}_n&=& \sum_{\{i , j , k , l\}}  
c(i,j,k,l)  \,
K^\mu F(i,j,k,l) \,
\nonumber \\
&=& \, -2\epsilon   \sum_{i=1}^n \sum_{k=i+2}^{i+n-3} 
\cE(i,k)
\big[  x_{i-1}^{\mu}\,
x_{ik}^2\,J(x_{ik}^2,x_{i-1\, k}^2)- x_{i}^{\mu}\,
x_{i-1\,k}^2\,J(x_{i k}^2,x_{i-1\, k}^2)\big] 
\nonumber \\
\ && \, -2 \epsilon   \sum_{i=1}^n
\Big[
\sum_{j=i+1}^{i+n-3}
-  
{c}(i,j,i-2,i-1) \Big]
x_{i-1}^{\mu}\, x_{ii-2}^2\,J(x_{ii-2}^2) 
\ , 
\label{eq:8}
\eeqa
where we have defined
\begin{align}
\label{Edef}
  \cE(i,k):= \sum_{j=k+1}^{i+n-2}
 {c}(i,k,j,i-1)-\sum_{j=i+1}^{k-1}
 {c}(i,j,k,i-1)\ ,
\end{align}
and  the formula \eqref{Edef} is only valid for $i<k$; if $i>k$, then
the variable $k$ appearing in the summation ranges of  \eqref{Edef} has
to be replaced by $k+n$.
Note that in \eqref{eq:8} all one-mass triangles are confined to the last line.
It is also important to mention that the second line in \eqref{eq:8} is of order $\epsilon^{-1}$, 
while the first  line is of order $\epsilon^0$.

Let us now discuss consequences of \eqref{eq:8} assuming the validity of \eqref{eq:2}  
up to and including terms of order $\mathcal{O}(\eps^0)$.

First, we observe that the term in square brackets in the last line of \eqref{eq:8}
is nothing but the particular combination  of two-mass hard and one-mass box coefficients  
discovered in \cite{dissolving}%
\footnote{Incidentally, this is the same combination that appears  in the  BCF recursion relation \cite{bcf}.}
\beq
\label{abc}
\sum_{j=i+1}^{i+n-3} 
{c} (i-2, i-1, i, j)  \ = \ 2 \, \mathcal{A}^{\rm tree}_n , \ i = 1, \ldots ,n  \  . 
\eeq
We find it interesting that the  combination of infrared consistency conditions introduced in \cite{dissolving} 
and derived in \cite{cahk}  emerges  naturally from our conformal equations. 
Hence, the last line of \eqref{eq:8} captures entirely the dual conformal anomaly \eqref{eq:2} of \cite{dhks}.

Second, by comparing \eqref{eq:8} and \eqref{eq:2} we find that the second line of  \eqref{eq:8} must vanish.
Since each coefficient $\cE(i,k)$ is multiplied by an independent combination of two-mass triangles they must
vanish individually, leading to the following set of equations:
\beq
\cE(i,k) \ = \ 0 \ ,  \qquad  i=1,
\ldots, n\ ,\ \ k=i+2, \ldots , i+n-3 \ . 
\label{gennconfeq2}
\eeq
These equations are among the most important new results of this paper, because they
give new constraints on the box supercoefficients arising from dual conformal invariance. 
In Section \ref{analysisconfeqns} we will discuss in more detail consequences of these additional conformal equations, 
where we will also prove that these lead to 
$n(n-4)$ independent linear relations among the one-loop supercoefficients. 
Notice that
assuming the conformal anomaly equation \eqref{eq:2} we also have found an alternative proof of
\eqref{abc}.

For later convenience we extend the definition of $\cE(i,k)$ to include also $\cE(i,i-2)$ and $\cE(i-1,i)$,
\beq
\label{eii-2}
\cE(i, i-2) \, = \, - \cE(i-1,i) \, := \, - \sum_{j=i+1}^{i+n-3} 
{c} (i-2, i-1, i, j) 
\ ,
\eeq
which allows us to rewrite \eqref{abc} as
\beq
\label{gennconfeq1}
\cE(i,i-2) \ = \ - \cE(i-1,i) \ = \ - 2 \, \mathcal{A}^{\rm tree}_n \ , \ i = 1, \ldots ,n \   .
\eeq

\subsection{Examples of applications of the conformal equations}

In order to make more concrete the discussion presented in the previous section, we would now like to apply  \eqref{gennconfeq1} and \eqref{gennconfeq2} 
 to the case of four-, five- and six-point superamplitudes. 

\subsubsection{Four- and five-point amplitudes} 
\label{sec:four-five-point}

The four- and five-point cases are simple enough that we can address them directly,   without passing through the 
$n$-point conformal equations derived in \eqref{gennconfeq2}, \eqref{gennconfeq1}. 

At four points, the only box function that can appear is the massless box. 
The one-loop superamplitude can then be written as  
\begin{align}
\label{ffour}
  \mathcal{A}_{4}^{\mathrm{1-loop}} \ =  \  
 c^\mathrm{0m}  \, F^\mathrm{0m} \ .
\end{align}
The action of a special conformal transformation on the superamplitude is expected to produce the anomaly
 \eqref{eq:2}. Explicitly, we expect to find 
\begin{align}
\hat{K}^\mu   {\mathcal{A}}_{4}^{\mathrm{1-loop}} \,  = \, 4\epsilon\, 
{\mathcal{A}}_{4}^{\mathrm{tree}} \Big[
(x_{2}^\mu+x_{4}^\mu)
  x_{1 3}^2 J(x_{1 3}^2)\, +\, 
  (x_{1}^\mu+x_{3}^\mu) x_{2 4}^2 J(x_{2  4}^2)
  \Big] \ . 
\end{align}
By acting directly on \eqref{ffour}, we obtain 
\begin{align}
\label{2ffour} 
\hat{K}^\mu   {\mathcal{A}}_{4}^{\mathrm{1-loop}} \,  = \, 
  {c}^\mathrm{0m}  \, K^\mu F^\mathrm{0m}\ . 
\end{align}
In writing \eqref{2ffour}, we have used  
the invariance of the box coefficients under the action of $\hat{K}_\mu$ \cite{bhtrec,dhks}.

The conformal transformation of
the zero-mass box function is given in \eqref{cieq}, and inserting
this in \eqref{2ffour} we find that the one-loop four-point amplitude is 
dual conformally invariant as long as 
$
c^\mathrm{0m}=2\, {\mathcal{A}}_{4}^{\mathrm{tree}}$, i.e.
  \begin{align}\label{eq:4}
  \mathcal{A}_{4}^{\mathrm{1-loop}} \ =  \ 
  2\,   \mathcal{A}_{4}^{\mathrm{tree}} \, F^\mathrm{0m} \ .
\end{align}

At five points, only one-mass box functions can appear, therefore the generic one-loop superamplitude can be written as 
\begin{align}
\label{gen5pt} 
    \mathcal{A}_{5}^{\mathrm{1-loop}} \ =  \  
    \sum_{i=1}^{5}
 c^\mathrm{1m}_i  \, F^\mathrm{1m}_i \ .
 \end{align}
Applying $\hat{K}^\mu$ to the left hand  side of \eqref{gen5pt} and using the expression for the anomaly in \eqref{eq:2}, 
one expects to find
\begin{align}
 \hat{K}^\mu {\mathcal{A}}_{5}^{\mathrm{1-loop}} \ =  \  4 \epsilon\,
   {\mathcal{A}}_{5}^{\mathrm{tree}} 
x_{1}^\mu x_{2 5}^2 \,J(x_{2 5}^2)\,  + \, \mathrm{cyclic}\ .
  \label{eq:5}
\end{align}
Here `$+ \, \mathrm{cyclic}$' indicates the addition of four terms
obtained by rewriting the labels as
$1\rightarrow 2\rightarrow  3\rightarrow 4\rightarrow 5\rightarrow 1$.
Now applying $\hat{K}^\mu$ to the right hand  side of \eqref{gen5pt} gives 
\begin{align}
\nonumber
  \sum_{i=1}^5
  c^\mathrm{1m}_i  \, \hat{K}^\mu F^\mathrm{1m}_i&= 
-2\eps \,   c^\mathrm{1m}_1
 \Big\{
- x_1^\mu x_{24}^2 J(x_{24}^2 , x_{41}^2)  + x_2^\mu \Big[ x_{41}^2 J (x_{24}^2, x_{41}^2 ) - x_{13}^2 J (x_{13}^2) \Big] 
\\ \nonumber 
&\hspace{0.5cm} + x_3^\mu \Big[ x_{41}^2 J (x_{13}^2, x_{41}^2 ) - x_{24}^2 J (x_{24}^2) \Big]  
- x_4^\mu x_{13}^2 J(x_{13}^2 , x_{41}^2)\Big\} + \mathrm{cyclic}\, , 
\\
\label{ff1}
\end{align}
or, after rearranging terms, 
\begin{align}
\nonumber
  \sum_{i=1}^5
  c^\mathrm{1m}_i  \, \hat{K}^\mu F^\mathrm{1m}_i 
&=-2\epsilon \,x_1^\mu \Big\{ ( c_4^\mathrm{1m}- c_1^\mathrm{1m})\, 
x_{24}^2 J(x_{24}^2 , x_{41}^2) \, +\, 
( c_5^\mathrm{1m}- c_3^\mathrm{1m}) 
\, 
x_{35}^2 J(x_{13}^2 , x_{35}^2)\\&
\hspace{0.5cm} -( c_4^\mathrm{1m}+ c_5^\mathrm{1m}) \, x_{25}^2
J(x_{25}^2)\Bigr\} \, + \, 
\mathrm{cyclic}
\ . 
\label{eq:6}
\end{align}
Equation \eqref{ff1}  comes directly  from \eqref{cieq}, whereas in 
\eqref{eq:6} we have collected  all terms proportional
to $x_1^\mu$.

On comparing \eqref{eq:6} with \eqref{eq:5}  we learn  that,
for dual conformal invariance to hold, the first line of \eqref{eq:6} must
vanish, implying that all the one-mass coefficients are equal;  furthermore,  from the
second line of  \eqref{eq:6}, we also obtain 
\begin{align}
   c_i^\mathrm{1m}= {\mathcal{A}}_{5}^{\mathrm{tree}}\ , \qquad
   i=1, \ldots, 5 \ .
\end{align}
We thus find that at five points,  
\begin{align}\label{eq:7}
    \mathcal{A}_{5}^{\mathrm{1-loop}} \, =  \,  
\mathcal{A}_{5}^{\mathrm{tree}}  \, \sum_{i=1}^5 \,F^\mathrm{1m}_i 
\ .
\end{align}
Note that since we consider the full superamplitude, the above results \eqref{eq:4} and \eqref{eq:7}
apply to all  four- and five-point superamplitudes in $\mathcal{N}\! =\!4$
SYM. Of course, all four-point amplitudes are MHV, and all five-point
amplitudes are either MHV or anti-MHV,  and the above
results \eqref{eq:4} and  \eqref{eq:7} are well-known. 
We also notice that these results can also be obtained by just imposing the known infrared consistency 
conditions on the one-loop superamplitudes.

\subsubsection{Six-point amplitudes}
\label{sec:six-point-amplitudes}

At six points, there are a total of six possible two-mass hard
coefficients, $c^\mathrm{2mh}(i,i+4)$, three two-mass easy
coefficients $c^\mathrm{2me}(i,i+3)$, six one-mass coefficients
$c^\mathrm{1m}(i)$ and no three-mass coefficients. 

We will now use the conformal equations \eqref{gennconfeq2}, \eqref{gennconfeq1}
to find relations among these coefficients. At the six-point level, these equations become
\begin{align}\label{eq:15}
  \cE(i,i+1)&\equiv c^\mathrm{1m}(i-1)+ c^\mathrm{2mh}(i-1,i+3)+
  c^\mathrm{1m}(i-2) &=& \ \, 2 \, \cA^{\mathrm{tree}}_6
  \, , 
  \\
  \cE(i,i+2)&\equiv c^\mathrm{2mh}(i-2,i+2)- c^\mathrm{1m}(i-1) +
  c^\mathrm{2me}(i-1,i+2) &=& \ \, 0 \, , 
  \label{bbbc} \\
\cE(i,i+3)&\equiv c^\mathrm{1m}(i+3)- c^\mathrm{2mh}(i-1,i+3)-
  c^\mathrm{2me}(i-1,i+2) &=& \ \, 0\, .
  \label{cccd}
\end{align}
The second and third equations imply that there are in fact only three
independent two-mass hard coefficients, since
\beqa
  c^\mathrm{2mh}(4,1)&=&c^\mathrm{2mh}(1,5)\ , 
  \nonumber \\
  c^\mathrm{2mh}(5,2)&=&c^\mathrm{2mh}(2,6)\ , 
  \label{eq:16}\\
  c^\mathrm{2mh}(6,3)&=&c^\mathrm{2mh}(3,1)
  \ .\nonumber
\eeqa
We can choose for instance $c^\mathrm{2mh}(1,5)$,
$c^\mathrm{2mh}(2,6)$, and $c^\mathrm{2mh}(3,1)$.

Equations \eqref{bbbc} and \eqref{cccd} can also be used to show that all 
two-mass easy coefficients are equal and so there is only one
independent two-mass easy coefficient (e.g.~$c^\mathrm{2me}(1,4)$).
As for the one-mass coefficients,  these are determined by the
remaining independent two-mass coefficients:
\begin{align}
  c^\mathrm{1m}(1)&=c^\mathrm{1m}(4)=c^\mathrm{2mh}(3,1)+c^\mathrm{2me}(1,4)\ , 
  \nonumber\\
  c^\mathrm{1m}(2)&=c^\mathrm{1m}(5)=c^\mathrm{2mh}(1,5)+c^\mathrm{2me}(1,4)\ , 
  \label{eq:17}
  \\
  c^\mathrm{1m}(3)&=c^\mathrm{1m}(6)=c^\mathrm{2mh}(2,6)+c^\mathrm{2me}(1,4)
  \  .
  \nonumber 
\end{align}
Finally the first equation \eqref{eq:15} gives one of the remaining
four independent coefficients in terms of the tree-level superamplitude $\cA^{\mathrm{tree}}_6$. 
For example,
\begin{align}
\label{van}
  c^\mathrm{2me}(1,4) =  \cA^{\mathrm{tree}}_6 - {1\over 2} \left[
    c^\mathrm{2mh}(1,5)+c^\mathrm{2mh}(2,6)+c^\mathrm{2mh}(3,1)
  \right]\ .
\end{align}
Notice that these equations hold for arbitrary  six-point superamplitudes,
i.e.~they apply to both MHV and NMHV helicity assignments. 
The MHV case is instantly obtained from the previous relations by further setting 
$c^\mathrm{2mh}_\mathrm{MHV}=0$, 
so that 
$c^\mathrm{2me}_\mathrm{MHV}=\cA^\mathrm{tree}_\mathrm{MHV}$.
For the NMHV case, consider for instance the gluonic amplitudes calculated in \cite{fusing}. 
From \cite{fusing} we infer that the sum of coefficients  in square brackets in \eqref{van} 
is equal to $2\mathcal{A}^\mathrm{tree}$, and hence all the two-mass easy coefficients are zero, 
in agreement with \cite{fusing}.

The general MHV superamplitude is discussed further in Section \ref{sectionMHV}, while the $n$-point 
NMHV superamplitudes are discussed in 
Section \ref{sec:conf-invar-nmhv}.

\section{Analysis of the conformal equations}
\label{analysisconfeqns}

In this section we discuss the conformal equations derived earlier in \eqref{gennconfeq1} and \eqref{gennconfeq2}.
Firstly, we will show that the conformal equations imply the infrared consistency conditions at one loop. 
We will then move on to analyse what are the new constraints on supercoefficients arising from the 
conformal equation in addition to the infrared ones.

\subsection{Comparison with infrared consistency conditions}
\label{sec:comp-with-infr}

Here we compare the above equations from dual conformal invariance
with the infrared consistency conditions for one-loop scattering amplitudes. 
We will show that the infrared equations can be derived in a neat way from the conformal equations.
As anticipated earlier, in addition to these, the conformal equations also contain new relations  which further 
constrain the one-loop supercoefficients. In the last part of this section, we will present 
these relations explicitly. 

We start off by recalling that the infrared consistency conditions 
arise from the relation
\begin{equation}
\label{ircr} 
\mathcal{A}^{\mathrm{1-loop}}_n |_{\mathrm{IR}} \ = \ - r_{\Gamma} \, 
\mathcal{A}^{\mathrm{tree}}_n \sum_{i=1}^n 
\frac{(-x_{i i+2}^2)^{-\epsilon}}{\epsilon^2} \, .
\end{equation}
Now we wish to rewrite this equation as
\beq
\label{eq:IR}
{\mathcal{A}}^{\mathrm{1-loop}}_n|_{\mathrm{IR}}:=\  \, {r_{\Gamma} \over 2 \epsilon^2}   \sum_{i=1}^n \sum_{k=i+2}^{i+n-2} 
\cI(i,k) \
(-x_{ik}^2)^{-\epsilon} \ ,
\eeq
where obviously $\cI(i,k)=\cI(k,i)$.
By comparing \eqref{ircr} with \eqref{eq:IR} we obtain the following infrared consistency conditions:  
\begin{eqnarray}
  \cI(i,i+2) &= &\cI(i+2,i)\  = \  -2 \, \cA^{\rm tree}_n  \ , 
  \label{eq:9}
  \\  \cr
       \cI(i,k)&=& 0 \ , \quad
  k=i+3, \ldots ,  i-3  \ . 
\label{eq:100}
\end{eqnarray}

Now we would like to re-express  $ \cI(i,i+2)$ and 
 $\cI(i,k)$  in terms of the particular combinations of coefficients 
 $\cE(i,k)$ introduced earlier. In order to do so, we first observe that 
the conformal variation of the one-loop amplitude \eqref{eq:8} can be written more compactly as 
\beq
\hat{K}^{\mu}  {\mathcal{A}}^{\mathrm{1-loop}}_n\ =\ 
  \,-2\, \epsilon   \sum_{i=1}^n \sum_{k=i+2}^{i+n-2} 
\cE(i,k)
\big[  x_{i-1}^{\mu}\,
x_{ik}^2\,J(x_{ik}^2,x_{i-1\, k}^2)- x_{i}^{\mu}\,
x_{i-1\,k}^2\,J(x_{i k}^2,x_{i-1\, k}^2)\big]
\ ,  
\label{eq:888}
\eeq
where $\cE (i , i-2)$ is defined in \eqref{eii-2}.

Next, we use  the mapping between the
 conformal variations of the boxes with their infrared divergent
 parts discussed below \eqref{eq:1}. 
According to this mapping, in order to obtain the infrared divergent terms from the conformal 
anomalous terms,  every occurrence in the latter of $\epsilon \, x_i^\mu x_{jk}^2 J(x_{jk}^2, x_{ik}^2)$ 
is  mapped to a term  $-r_{\Gamma} (-x_{jk}^2)^{-\epsilon}/(4\epsilon^2)$ in the former.
Applying this map to \eqref{eq:888},  we arrive  at  the following structure for the 
infrared divergent terms of the amplitude:
\beq
\label{eq:irdiv}
{\mathcal{A}}^{\mathrm{1-loop}}_n|_{\mathrm{IR}} \,
= \, {r_{\Gamma} \over 2 \epsilon^2}   \sum_{i=1}^n \sum_{k=i+2}^{i+n-2} 
\cE(i,k)
\ \big[ 
(-x_{ik}^2)^{-\epsilon}  - 
(-x_{i-1\,k}^2)^{-\epsilon} \big] 
\ . 
\eeq
Symmetrising the right hand side of \eqref{eq:irdiv} in $i$ and $k$, we arrive at
\beq
{r_{\Gamma} \over 4 \epsilon^2}   \sum_{i=1}^n \sum_{k=i+2}^{i+n-2} 
\Big\{ \cE(i,k) \big[ 
(-x_{ik}^2)^{-\epsilon}  - 
(-x_{i-1\,k}^2)^{-\epsilon} \big] + \cE(k,i)
\big[ 
(-x_{ik}^2)^{-\epsilon}  - 
(-x_{i\,k-1}^2)^{-\epsilon} \big]\Big\} \ ,
\eeq
and finally we manipulate this expression by
collecting together all terms proportional to $(-x_{ik}^2)^{-\epsilon}$ by shifting
indices and changing summation boundaries appropriately. 
Thus we find
\beqa
{\mathcal{A}}^{\mathrm{1-loop}}_n|_{\mathrm{IR}}  &= &  {r_{\Gamma} \over 4 \epsilon^2}   \sum_{i=1}^n \sum_{k=i+3}^{i+n-3} 
\big[ \cE(i,k) +\cE(k,i)-\cE(i+1,k)-\cE(k+1,i)\big]
(-x_{ik}^2)^{-\epsilon} \nonumber\\
& +& {r_{\Gamma} \over  2\epsilon^2}   \sum_{i=1}^n
\big[ \cE(i,i+2) +\cE(i+2,i)-\cE(i+3,i)\big]
(-x_{i\,i+2}^2)^{-\epsilon} 
\ .
\label{eq:11}
\eeqa
Note that special care has to be taken at the boundaries of the summation for coefficients of $(-x_{i\,i+2}^2)^{-\epsilon}$ which leads to the second line
in \eqref{eq:11}. 

Comparing \eqref{eq:11} to \eqref{eq:IR} we immediately get
\beqa
  \cI(i,i+2) & = & \cE(i,i+2)+\cE(i+2,i) -\cE(i+3,i) \ , \label{eq:12}
   \\ \nonumber 
  \cr
  \cI(i,k) & = &  \cE(i,k)+\cE(k,i) - \cE(i+1,k)-\cE(k+1,i)\ ,\quad k=i+3, \ldots , 
   i+n-3 \ . \nonumber \\ \nonumber
\eeqa
Finally, by plugging  the conformal equations \eqref{gennconfeq2} and 
\eqref{gennconfeq1} into \eqref{eq:12} one obtains  
 the infrared consistency conditions \eqref{eq:9} and \eqref{eq:100}.

We have mentioned earlier that the  particular combination of infrared equations  
introduced in \cite{dissolving} and proven in \cite{cahk} emerges naturally from our  conformal equations, 
namely it appears in  \eqref{gennconfeq1}. 
Here we would like to add that  this combination   arises from considering the
sum $ \sum_{k=i+2}^{i+n-2} \cI(i,k)$ of infrared equations. 
 %
Indeed, using the $2n$ algebraic identities
\begin{align}
\label{411}
  \sum_{k=i+1}^{i+n-2} \cE(i,k) &= 0
  \ , \\
  \label{412}
  \cE(i,i+1)&=-\cE(i+1,i+n-1)
\ , 
\end{align}
which can easily be checked by using the definition of $\cE(i, k)$ given in \eqref{Edef}, 
one obtains  
\begin{align} \label{eq:13}
  \sum_{k=i+2}^{i+n-2} \cI(i,k)\ = \ 2\, \cE(i+1,i-1)\ =\ -4\,  \cA_{\rm tree}
  \ , 
\end{align}
which is the combination of \cite{dissolving}.

Finally, one may want to disentangle the genuinely new conformal equations from the infrared consistency conditions. This can be achieved in the following way.

Firstly,  we consider the following linear combinations of the dual conformal
equations \eqref{gennconfeq1} and \eqref{gennconfeq2}: 
\begin{align}
\cF(i,k)&\equiv  \sum_{l=i+1}^{k-1} \cE(i,l) \equiv \sum_{l=i+1}^{k-1}
  \sum_{m=k}^{i+n-2} c(i,l,m,i-1)  
  \nonumber \\&\equiv\ 
  c^\mathrm{1m}(i-2)+c^\mathrm{2me}(i-1,k-1)+{\rm 
   ( 2mh,\ 3m \ coefficients)}\nonumber\\
 &=\  2 \,\cA^{\mathrm{tree}}_n 
 \  , 
 \label{eq:14}
\end{align}
where $k=i+2, \ldots ,  i+n-2$.  
The last equality in \eqref{eq:14}  follows from the conformal equations \eqref{gennconfeq1} and \eqref{gennconfeq2}.

On examination of the explicit expression in terms of box functions it
is clear that
\begin{align}
  \cF(i,i+2) \ \equiv \ \cF(i+1,i-1)\ .
\end{align}
Other than this all equations are independent, and since there are
$n(n-4)$ equations, the $\cF$ equations are completely
equivalent to the original $\cE$ equations.
 
Now we consider the antisymmetric combination $\cF(i,k)-\cF(k,i)$, and notice that it is
independent of the two-mass easy box coefficients since
$c^\mathrm{2me}(i-1,k-1)=c^\mathrm{2me}(k-1,i-1)$. We can further find combinations
which are completely independent of  two-mass easy as well as of the the one-mass box coefficients. These are given by 
the following expression, 
\beq
  \cG(i,k)\, := \, 
  \cF(i,k)-\cF(i+1,k)-\cF(i,k+1)+ \cF(i+1,k+1)\  -\ 
   ( i  \leftrightarrow k) 
  \ , 
  \eeq
where $k= i+3 , \ldots ,  i-3 $.
Since $\cG(i,k)=-\cG(k,i)$,  and for each $i=1,\ldots n$, the index $k$ takes on
$n-5$ values, we have $n(n-5)/2$ such equations. These  are  genuinely new relations  
from dual conformal invariance, i.e.~ they do not already arise from 
infrared consistency conditions alone.

\subsection{Solution to the conformal equations}

We begin by  discussing how many linearly independent relations arise from the conformal equations. 
We have $n(n-4)$ equations from  \eqref{gennconfeq2}, to which we have to add the $2n$  anomaly conditions 
\eqref{gennconfeq1}, which in total give  $n(n-2)$ conformal equations. 
Now we have found    $2n$ algebraic relations in \eqref{411} and \eqref{412}. 
This leaves exactly  $n(n-4)$ linearly independent conformal equations.%
\footnote{We have checked numerically up to $n=14$ that the remaining equations are indeed independent. }

An independent set of equations is given by considering the quantities 
$ \cE(i,k)$ for $k=i+2, \ldots ,  i+n-3$,  
which satisfy the relations 
\begin{align}
 \sum_{i=1}^n
  \sum_{k=i+2}^{i+n-3} \cE(i,k)\, = \, 0 \ . 
\end{align}
In addition, we include a  single equation giving the conformal anomaly,
\eqref{gennconfeq1}.

The next question we would like to address is how can we use the conformal equations to  solve 
for particular sets of  coefficients of the one-loop amplitudes. 
We have addressed this issue by studying  with an algebraic manipulation program 
the system of linear equations \eqref{gennconfeq2} and \eqref{gennconfeq1}, and solving for different choices of coefficients. 
If we define the particular (anti-)symmetric combinations  of two-mass hard coefficients
\beqa
c^{\mathrm{2mh}}_a (i, j) &:=&  c^\mathrm{2mh} (i, j) \, -\, 
c^\mathrm{2mh}(j-1,i+1)\ , 
\\ 
 c^{\mathrm{2mh}}_s(i,j)& := &  c^\mathrm{2mh}(i,j)\, +\, 
c^\mathrm{2mh}(j-1,i+1)\ , 
\eeqa
then the conformal equations can be generally recast as 
the following $n(n-4)$ equations
\beqa
  c^\mathrm{1m} (i)  &=&f^\mathrm{1m}_i  (  c^\mathrm{2mh}_s,    c^\mathrm{3m}) \ ,
\\
  c^\mathrm{2me} (i, j)  &=&  f^\mathrm{2me}_{ij}(  c^\mathrm{2mh}_s,    c^\mathrm{3m}) \ ,
\\
  c^{\mathrm{2mh}}_a (i, j)  &= &f^{\mathrm{2mh}; a}_{ij} (  c^{\mathrm{2mh}}_s,    c^\mathrm{3m}) \ ,
\eeqa
thus enabling us to solve for the $n(n-5)/2$ antisymmetric combination
of two-mass hard coefficients, as well as for the $n$ one-mass and $n(n-5)/2$ two-mass easy coefficients  
in terms of the symmetric combination of
two-mass hard coefficients, together with three-mass coefficients
only. The precise functions  in the equations above appear  to be quite complicated but we
have explicitly checked that this rearrangement is possible up to $n=12$. 

One comment is in order here. 
It is known that for $n$ odd, one may use the infrared equations to solve for all
one-mass and two-mass box coefficients in terms of two-mass hard and
three-mass coefficients \cite{Bern:2004bt}. 
For $n$ even, this is no longer possible since one of the infrared consistency equations
can no longer be used to this purpose. 
In our discussion before, we find that the cases of  $n$ even and odd are not distinguished.%
\footnote{Another possible approach would have been to solve for the two-mass hard coefficients and the one-mass coefficients in terms of the two-mass easy and the three-mass ones. It turns out that  one can do this when $n$ is not divisible by $3$.  }

Summarising,  we have shown that, after imposing dual conformal invariance,  the coefficients 
$c^\mathrm{1m}, c^\mathrm{2me}$ and $c^\mathrm{2mh}_a$ are
completely determined in terms of the remaining undetermined
coefficients $c^\mathrm{2mh}_s$ and $c^\mathrm{3m}$.

\section{The anomaly equations for MHV and NMHV superamplitudes}

The anomaly equations \eqref{gennconfeq2} and \eqref{gennconfeq1} found above give relations 
between box coefficients  of  generic  superamplitudes. 
In this section we will specialise to  particular cases, namely to the MHV, anti-MHV and NMHV superamplitudes, 
in order to provide explicit checks  of  these equations.

\subsection{MHV (and anti-MHV) amplitudes}
\label{sectionMHV}

MHV superamplitudes (and hence also the corresponding box
supercoefficients) have a total of eight $\eta$'s. 
Calculating these using unitarity cuts one quickly sees that two tree-level three-point anti-MHV
amplitudes are required at two of the corners and two tree-level MHV amplitudes
at the other two corners. Since anti-MHV amplitudes can not be
adjacent, the three-point anti-MHV amplitudes must lie at opposite
corners of the box and hence only two-mass easy and one-mass box
coefficients are non-vanishing.

Let us now consider \eqref{gennconfeq2}. For the MHV case this becomes, after shifting $i$ to $i+1$ for convenience,
\begin{align}
 \cE (i+1, k) \, = \, c^\mathrm{2me} (i, k) -   c^\mathrm{2me}(i, k-1)\, = \, 0
 \ , 
\end{align}
with $k = i+3 , \ldots , i+n-2$.
Considering this relation for all allowed values of $k$ one obtains that 
\begin{align}
  c^\mathrm{1m}(i)\, =\, c^\mathrm{2me}(i,k)  \, = \, c^\mathrm{1m}(k) \ ,   
\end{align}
where in the last step we have used $ c^\mathrm{2me}(i,k)  =  c^\mathrm{2me}(k,i) $. In other words, all the one-mass and two-mass easy coefficients are equal to each other. To determine their common value we use the anomaly equation 
\eqref{gennconfeq1}, which becomes 
\beq
\cE(i , i-2) =-\big ( c^\mathrm{1m}(i-2) + c^\mathrm{1m}(i-3)\big)  \, = \, -2 \, \cA^\mathrm{tree}_n
\ , 
\eeq
from which we infer  that all the supercoefficients are equal to $\cA^\mathrm{tree}_n$.

The four- and five-point cases discussed in
section \ref{sec:four-five-point} are special cases of this since one
can 
only have MHV or anti-MHV amplitudes in these cases.

\subsection{NMHV amplitudes}
\label{sec:conf-invar-nmhv}

The infinite sequence of one-loop NMHV gluon scattering amplitudes has been calculated in  \cite{Bern:2004bt}. 
Here we will consider the  NMHV superamplitudes, which have been calculated recently 
in \cite{dhksgen} using supersymmetric generalised unitarity. 
There the coefficients are all expressed in terms of the 
quantities  $ R_{rst}$, defined as
\begin{align}\label{eq:18}
   R_{rst} :=  \cA_{n, \mathrm{MHV}}^\mathrm{tree} \, 
   \frac{\vev{s-1\,s}\vev{t-1\,t}\  \delta^{(4)}\bigl(\Xi_{rst}\bigr)}{x_{st}^2\vev{r|x_{rt}x_{ts}|s-1}\vev{r|x_{rt}x_{ts}|s}
\vev{r|x_{rs}x_{st}|t-1}\vev{r|x_{rs}x_{st}|t}}\, , 
\end{align}
where
\beq
\label{Xifunction}
\Xi_{r st} \, := \, 
\langle  r | \left[ x_{rs} x_{st} \sum_{k=t}^{r-1} |k\rangle \eta_k+ x_{rt} x_{ts} \sum_{k=s}^{r-1} | k \rangle \eta_k
\right]
\ . 
\eeq
Notice that compared to \cite{dhks} we have included for convenience 
a factor of $ \cA_{n, \mathrm{MHV}}^\mathrm{tree}$ into the definition of $R_{rst}$.  
With this definition, the $R_{rst}$  are directly equal to the 
three-mass coefficients 
$ c^{\rm 3m}(r,s,t)$ whenever $r,s,t$ lie in the appropriate range,
$r+2<s<t-1<r+n-2$. In fact, 
all box coefficients can be written in terms of $R_{rst}$ as follows \cite{dhksgen}:  
\begin{align}
  c^{\rm 3m}(r,s,t)&=R_{rst} \ , 
  \label{eq:19}\\[10pt]
  c^{\rm 2mh}(r,t)&=R_{r,r+2,t} +R_{r+1,t,r}
  \ , \label{eq:20} \\[10pt]
  c^{\rm 2me}(r,s)&
=\sum_{u,v=s+1}^{r+n-1} R_{r,u,v}
     +\sum_{u,v=r+1}^{s-1} R_{s,u,v}  \ ,  
     \label{eq:21} \\[10pt]
     c^{\rm 1m}(r-2)&=c^{\rm 2me}(r,r-2)+R_{r-1,r+1,r-2}\nonumber\\[5pt]
     &=\sum_{u,v=r+1}^{r+n-3} R_{r-2,u,v}+R_{r-1,r+1,r-2}\ . \label{eq:22}
\end{align}
Note that $R_{ruv}$ is only defined for $ u\geq r+2$,  $v \geq u+2$
and 
$v \leq r+n-1$. We will define $R_{rst}=0$
outside these bounds. Thus $\sum_{u,v=s+1}^{r+n-1}:= \sum_{u=s+1}^{r+n-3}
\sum_{v=u+2}^{r+n-1} R_{r,u,v}$ and $\sum_{u,v=r+1}^{r+n-3}:=
\sum_{u=r+1}^{r+n-5} \sum_{v=u+2}^{r+n-3} R_{r-2,u,v}$ etc.

 A crucial relation between the $R$'s, which we will use in the following,  is  \cite{dhks}
\begin{equation}
\label{eq:23}
 R_{r+2,s,r+1}= R_{r,r+2,s}\ .
\end{equation}

Next, we would like to  prove that all  NMHV superamplitudes satisfy the
conformal equations \eqref{gennconfeq2}, \eqref{gennconfeq1}, and are thus
dual
conformal invariant. Specifying these equations to the NMHV case, and by inserting 
the expressions for the NMHV box
supercoefficients given in \eqref{eq:19}--\eqref{eq:21}  into the 
conformal equations, one obtains
\beqa
\nonumber
  \cE(i+1,i+2) &=& \, c^{\mathrm{1m}}(i) +c^{\mathrm{1m}}(i-1)+
  \sum_{j=i+4}^{i-2} c^{\mathrm{2mh}}(i,j)\\
  &=& \, \sum_{u,v=i+2}^{i+n-1} R_{i u v} + \sum_{u,v=i+1}^{i+n-2}
  R_{i-1 u v} \nonumber \\
&=& \, \, 2\,  \cA_{\mathrm{tree}}\ ,  \label{ireq}
  \eeqa
  as well as 
  \beqa
  \cE(i+1,k) &=& \, c^\mathrm{2me}(i,k) -  c^\mathrm{2me}(i,k-1) +
  \sum_{j=k+2}^{i+n-2} c^\mathrm{3m}(i,k,j) \nonumber  \\\nonumber 
   &-&\!\!\sum_{j=i+3}^{k-2} c^\mathrm{3m}(i,j,k) +
  c^\mathrm{2mh}(i-1,k)-c^\mathrm{2mh}(i,k) \nonumber \\
&=&\, \sum_{u,v=i+1}^{k-1}R_{k u v} - \sum_{u,v=i+1}^{k-2}R_{k-1 u v} -
\sum_{u=i+2}^{k-2} R_{iuk} \nonumber \\
&=&\, \left(\sum_{u,v=i+1}^{k-1} R_{kuv} - \sum_{u,v=i+2}^k R_{iuv}\right) - \left(
  \sum_{u,v=i+1}^{k-2} R_{k-1 uv} - \sum_{u,v=i+2}^{k-1} R_{iuv}
\right)\nonumber \\
&=&\, 0  \ , 
\label{confNMHV}
\eeqa
 where $k=i+3,  \ldots , i+n-2$. We have  also used the identity \eqref{eq:23} to
cancel some terms. Note that although the first equality
of \eqref{confNMHV} is 
strictly true only for $k=i+4, \ldots ,  i+n-3$ the second and third
equalities come out correct for the two cases $k=i+3$ and $k=i+n-2$ also.
In the sums over $u,v$ it is always assumed that $u\leq v-2$.

Since the tree-level 
NMHV superamplitude is given by \cite{dhks,dhksgen,Drummond:2008cr}
\begin{align}\label{eq:26}
  \cA^{\mathrm{tree}} =   \sum_{u,v=i+2}^{i+n-1} R_{i u v}
  \ , 
\end{align}
which is valid for any value of $i$, \eqref{ireq} is
clearly true. 
As for the relations in \eqref{confNMHV}, they are satisfied if the following relation holds: 
\begin{equation}
\label{eq:24}
   \sum_{u,v=r+2}^{s} R_{r u v} =
\sum_{u,v=r+1}^{s-1} R_{s u v} \ ,  \qquad \qquad r+5\leq s \leq r+n-1 \ . 
\end{equation}
Equation \eqref{eq:24}  is a new identity for the $R_{rst}$, and the NMHV superamplitude
is dual conformal invariant only  
if this new identity is true. In Appendix  \ref{appA} we prove that this
new identity is indeed true, and hence we have proved that all NMHV
superamplitudes are dual  conformal invariant.

\newpage

\section*{Acknowledgements}

It is a pleasure to thank   Bill Spence, Mark Spradlin, Cristian Vergu and Anastasia Volovich  for   interesting discussions.
PH and GT would like to thank the Physics Department at Brown University for their warm hospitality during the last stages of this project.  We would also like to thank the organisers of the ``International Workshop on Gauge and String Amplitudes" held at Durham University, March 30--April 3 2009,  for giving us the opportunity to present the results of this paper. 
This work was supported by the STFC under a Rolling Grant  ST/G000565/1.
The work of PH is supported by an EPSRC Standard Research Grant EP/C544250/1.
GT is supported by an EPSRC Advanced Research Fellowship EP/C544242/1
and by an EPSRC Standard Research Grant EP/C544250/1.

\appendix

\section{Dual conformal invariance of all NMHV superamplitudes}
\label{appA}

In Section \ref{sec:conf-invar-nmhv} we have reduced the problem of
proving dual conformal invariance of the NMHV superamplitude to that
of proving the new identity \eqref{eq:24}. In the following we give the proof
of this relation.

\begin{figure}[ht]
\begin{center}
\includegraphics[width=13cm]{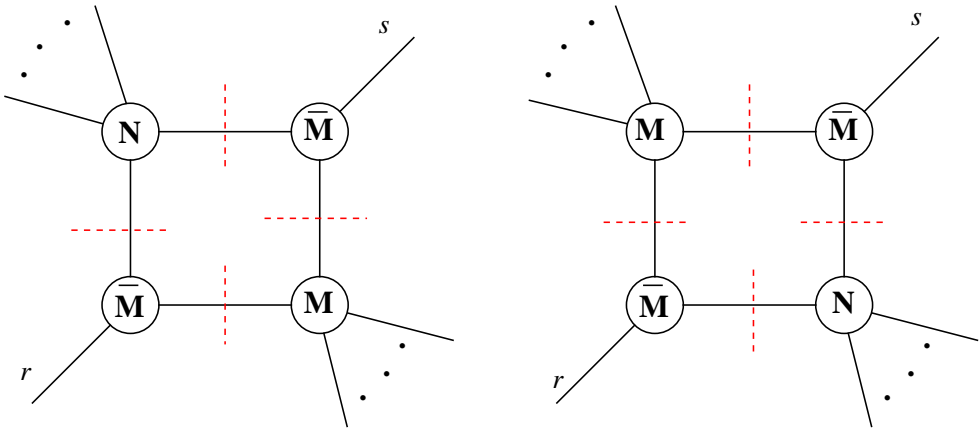}
\end{center}
\caption{\it  The two cut diagrams which contribute to the
  2me box function coefficient of the NMHV amplitude. Here in the corners
  $M$ represents a tree-level MHV amplitude, $\bar M$ the
  three-point anti-MHV amplitude and $N$ the tree-level NMHV amplitude. 
}
\label{c2mefig}
\end{figure}

 The two-mass easy box
coefficients  of the one-loop NMHV superamplitude are given by the sum of two supersymmetric quadruple cut diagrams \cite{dhksgen} represented in Figure \ref{c2mefig}. 
When computing these diagrams, one needs to insert the tree-level
NMHV superamplitude \eqref{eq:26} at the corner $N$ in Figure \ref{c2mefig}. 
There are then two natural choices for the 
label $i$ in \eqref{eq:26}, corresponding to either of the two loop
momenta leaving the corner $N$. These two choices give two different
expressions for the cut diagram. 
Consider for instance the first cut diagram.  It turns out that both sides
of the stronger identity \eqref{eq:24} precisely correspond to the two
different forms for this quadruple cut diagram. 
This observation  proves the identity 
\eqref{eq:24}, and hence completes the proof
that all NMHV superamplitudes are dual conformally covariant.

The identity \eqref{eq:24} is very similar to a weaker identity, 
\begin{align}\label{eq:25}
\sum_{u,v=r+2}^{s}  R_{ruv} + \sum_{u,v=s+2}^{r+n} R_{suv} = 
\sum_{u,v=s+1}^{r+n-1} R_{ruv} + \sum_{u,v=r+1}^{s-1}  R_{suv}
\ ,   
\end{align}
where without loss of generality we restrict $r+5 \leq s \leq
r+n-1$.
This identity was conjectured in the bosonic case
in \cite{Bern:2004bt}, and proved in the supersymmetric case in \cite{dhksgen}.  
The proof used a similar argument to that outlined above but
applied to the two-mass easy box coefficient itself (i.e.~the sum of the
two cut diagrams). 
The two sides of the weaker identity \eqref{eq:25} then give two different
representations of the two-mass easy box coefficient
$c^{\rm 2me}(r,s)$  \cite{dhksgen}.

Interestingly, it turns out that the weaker relation \eqref{eq:25} is completely
equivalent to the
stronger one \eqref{eq:24} whenever $s$ is restricted to $r+n-4 \leq   s \leq
r+n-1$. To see this, note that the second sum on the left-hand side and the
first sum on the right-hand side of \eqref{eq:25} are empty for $r+n-3 \leq   s \leq
r+n-1$ (recall that $u\leq v-2$). For the case $s=r+n-4$, on the other hand,
there is only one 
term in the second sum on the left-hand side and the
first sum on the right-hand side of \eqref{eq:25}, and  using the known identity
$R_{r-4\,r-2\,r}=R_{r\,r-3\,r-1}$, obtained from two applications
of \eqref{eq:23}, we again find the equivalence of \eqref{eq:24}
and \eqref{eq:25}. 
Therefore, \eqref{eq:24} is a new identity only when 
$r+5 \leq s \leq r+n-5$, which requires $n \geq 10$.

Summarising, \eqref{eq:25} is sufficient to prove dual 
conformal invariance of the NMHV amplitude as long as $n\leq9$, 
as was done in \cite{dhksgen}.
The new  identity we presented in \eqref{eq:24} is required 
for a proof valid for arbitrary $n$.

We conclude this appendix by mentioning an alternative proof of \eqref{eq:24}.
As a straightforward consequence
of the form of the tree-level NMHV superamplitude \eqref{eq:26},
we obtain
\begin{align}
\label{4.19}
\sum_{u,v=r+2}^{s} R_{r u v} \, =\, 
\sum_{u,v=r+1}^{s-1} R_{s u v}
\ . 
\end{align}
in the special case when the number of scattered particles  is $s-r+1$.
This is just the identity  \eqref{eq:24} that we wish to prove, for this
restriction on the number of particles.
If we now assume that this identity
does not require (super-)momentum 
conservation%
\footnote{This is claimed in \cite{dhksgen} as long as we take the
identity precisely as written, and do not use (super)-momentum
conservation to write, for example, $R_{s u v}=R_{s-n\, u v}$ etc.},
then we can use exactly 
the same identity for an arbitrary  number of external particles. 
In other words, we can relax the constraint on $s$,  thus obtaining the
identity   \eqref{eq:24} for arbitrary $r,s$, as required.

\section{Dual conformal combinations of box functions}

The equations \eqref{gennconfeq2} and \eqref{gennconfeq1}   can be
simply adapted for a closely related purpose, namely that of  finding
conformally invariant linear combinations of box functions with
constant coefficients. The steps needed to solve this problem 
are very similar to those leading
to \eqref{gennconfeq2} and  \eqref{gennconfeq1}. There we consider a linear
combination of boxes (with non-constant coefficients) and seek
anomalous dual
conformal invariance.

Now we require instead that the coefficients are constant, and that there is no
anomaly. The resulting equations for the constant coefficients are
therefore just
\begin{eqnarray}
\cE(i,k) & = & 0 \ ,  \qquad  i=1,
\ldots, n\ ,\ \ k=i+1, \ldots , i+n-2
\ . 
\end{eqnarray}
There are $n(n-4)$ such equations constraining $(1/2) n(n-5)(n-3)+n$
coefficients, leaving $(1/2) n (n-5 )^2$  dual conformal invariant
combinations.
This should be compared with finite combinations of boxes. These are
linear combinations of boxes whose constant coefficients satisfy
\begin{eqnarray}
       \cI(i,k)&=& 0 \ , \quad
  k=i+2, \ldots ,  i+n-2  \ . 
\label{eq:10}
\end{eqnarray}
giving $n(n-3)/2$ independent equations for $(1/2) n(n-5)(n-3)+n$
coefficients. There are therefore $n(n-5 ) (n-4 )/2$ finite
combinations of boxes.

For example at six points there are 6 finite combinations and 3 of
these are also
dual conformally invariant.
This differs from a statement
in \cite{dhksgen} that at six points all finite combinations of boxes
are conformally invariant. These six finite combinations are given by
\begin{equation}
  v_i=-2 F^\mathrm{1m}_{i}  - F^\mathrm{2me}_{i-1\, i+2}  -
  F^\mathrm{2me}_{i \,i+3} - F^\mathrm{2me}_{i+1\,i+4} + 2
  F^\mathrm{2mh}_{i\,i+4} +  
 2 F^\mathrm{2mh}_{i+1\,i-1} 
\ , 
\end{equation}
for $i=1,\ldots , 6$.
The 3 dual conformally
covariant combinations (which were given in \cite{dhksgen} 
and denoted there as  $v_{146}$, $v_{251}$ and $v_{362}$) are
\begin{align}
 v_{146}\ &=\  {1\over 4}\,  (v_0 + v_1 - v_2 + v_3 + v_4 - v_5)\ , \\ \nonumber
 v_{251}\ &=\  {1\over 4}\,  (-v_0 + v_1 + v_2 - v_3 + v_4 + v_5)\ , \\ \nonumber
 v_{362}\ &=\  {1\over 4}\,  (v_0 - v_1  + v_2 + v_3 - v_4 + v_5)\ .
\end{align}

At seven points there are 21 finite combinations of boxes and 14 of
these are conformally invariant. This agrees with the findings
 of   \cite{dhksgen}.

\vspace{1cm}


\end{document}